# Blue light-emitting diode based on ZnO

Atsushi TSUKAZAKI[1], Masashi KUBOTA[2], Akira OHTOMO[1], Takeyoshi ONUMA[2,3], Keita OHTANI[4], Hideo OHNO[4], Shigefusa F. CHICHIBU[2,3], and Masashi KAWASAKI[1,*,†]

[1]*Institute for Materials Research, Tohoku University, Sendai 980-8577, Japan*

[2]*Institute of Applied Physics and Graduate School of Pure and Applied Sciences, University of Tsukuba, Tsukuba 305-8573, Japan*

[3]*NICP, ERATO, Japan Science and Technology Agency, Kawaguchi 332-0012, Japan*

[4] *Laboratory for Nanoelectronics and Spintronics, Research Institute of Electrical Communication, Tohoku University, Sendai 980-8577, Japan*

A near-band-edge bluish electroluminescence (EL) band centered at around 440 nm was observed from ZnO *p-i-n* homojunction diodes through a semi-transparent electrode deposited on the *p*-type ZnO top layer. The EL peak energy coincided with the photoluminescence peak energy of an equivalent *p*-type ZnO layer, indicating that the electron injection from the *n*-type layer to the *p*-type layer dominates the current, giving rise to the radiative recombination in the *p*-type layer. The imbalance in charge injection is considered to originate from the lower majority carrier concentration in the *p*-type layer, which is one or two orders of magnitude lower than that in the *n*-type one. The current-voltage characteristics showed the presence of series resistance of several hundreds ohms, corresponding to the current spread resistance within the bottom *n*-type ZnO. The employment of conducting ZnO substrates may solve the latter problem.

*Also at Combinatorial Material Science and Technology (COMET), Tsukuba 305-0044, Japan
†E-mail address: kawasaki@imr.tohoku.ac.jp

The recent success of *p*-type doping in ZnO has opened a door to realize practical application of blue and ultraviolet (UV) light emitting diodes (LED), which can be an alternative to those based on III-nitrides[1,2]. Comparing with nitrides, ZnO has significant advantages from following reasons; (1) abundant mineral sources of Zn, (2) commercial availability of large area and high quality single crystal substrates[3-5] and (3) large exciton binding energy particularly advantageous for making highly efficient lasers.[6-10] However, the device structure of the LEDs reported previously had not been optimized in such ways as thick and opaque top electrode and asymmetric bottom electrode structure.[1] The previous device had square-shaped mesa structure with a glide type bottom electrode, giving rise to inhomogeneous current path within the *n*-type bottom current spread layer. These factors restricted us from observing electroluminescence (EL) spectrum intrinsic to the *p-i-n* junction; the detected light was limited to that coming from aside of the opaque top electrode close to the ion-etched walls of the mesa structure. Here, we have optimized the device structure to solve these problems by employing a semi-transparent top electrode and coaxial ring-shaped structure for the bottom electrode. The EL spectra were dominated by the near-band-edge emission in the *p*-type layer, which can be attributed to the donor-acceptor pair (DAP) luminescence related with doped nitrogen similar to that in Mg doped *p*-type GaN.[11] We identify the near-future challenges for improving the device performance based on the results.

The ZnO *p-i-n* homojunction structure was grown on lattice-matched (0001) ScAlMgO$_4$ (SCAM) substrates by laser molecular-beam epitaxy in $1\times10^{-6}$ Torr of oxygen. For nitrogen doping, a radio frequency radical source was used with the input power of 350 W and nitrogen flow of $5\times10^{-6}$ Torr as a background equivalent pressure. A semiconductor laser beam ($\lambda$ = 808 nm) was employed to heat the backside of the substrate susceptor. The device structure is schematically shown in Fig. 1. First, an 100-nm-thick ZnO buffer layer was deposited at 650 °C on the substrate and annealed *in-situ* at 1000 °C in 1mTorr of oxygen for 1 h.[12] All of the successive layers were grown in a layer-by-layer growth mode as verified by persistent intensity oscillation of reflection high-energy electron diffraction. As a bottom contact and current spread layer, a 400-nm-thick *n*-type ZnO doped with Ga was deposited, of which doping concentration was $2\times10^{18}$ cm$^{-3}$. After the undoped ZnO layer was deposited, *p*-type ZnO was grown by repeated temperature modulation (RTM) technique.[1] A key to realize *p*-type ZnO was revealed to keep both high nitrogen concentration (~ $10^{20}$ cm$^{-3}$) and low defect density. The former requires low temperature (400 $^{\circ}$C) deposition,[13] whereas the latter favors high temperature (1000 $^{\circ}$C) crystal growth as clearly revealed by positron annihilation and time-resolved photoluminescence experiments.[14] RTM technique effectively overcomes this dilemma. By repeating the growth of an approximately 15-nm-thick ZnO:N at low temperature followed by the high temperature annealing and subsequent growth of an 1-nm-thick ZnO at high temperature,[1] *p*-type layers having hole concentrations higher than $10^{16}$ cm$^{-3}$ were reproducibly obtained.

On the top of the *p*-type ZnO layer, semi-transparent Au (5 nm)/Ni (5 nm) was deposited by evaporation. The circular-shaped mesa with 300 μm in diameter was then fabricated by photolithography and Ar ion milling. Thick Au (100 nm)/Ni (10 nm) top contact (100 μm in diameter) was formed by the lift-off technique. Bottom electrode attached to the exposed *n*-type ZnO: Ga layer was made of Au (100 nm)/Ti (10 nm) with a ring shape, which was also fabricated by the lift-off technique. These Au/Ni and Au/Ti electrodes provide with good ohmic contacts for *p*-type and *n*-type ZnO layers, respectively. Figures 2(a) and 2(b) show photographs of the devices taken under illumination and with feeding a direct forward bias current of 5 mA in dark, respectively. The two contact probes attached on the top and bottom electrodes are visible in Fig. 2(a), and blue emission through the semi-transparent portion of the electrode is clearly seen in Fig. 2(b).

EL spectra of the *p-i-n* LED are shown in Fig. 3(a) as a function of forward current. The spectra are seen to exhibit three independent peaks at 395, 420 and 500 nm. Additionally, a weak shoulder is found in the spectra at around 610 nm. These structures can be explained as a convolution of light modulations due to the LED structure on the fundamental blue EL band centered at around 440 nm in the *p*-type ZnO layer, as follows. Photoluminescence (PL) and transmittance spectra of the identical *p*-type ZnO:N layer and a PL spectrum of the undoped ZnO are shown in Fig. 3(b). As shown, higher-energy cut-off wavelength of the EL spectra (390 nm, indicated by thick arrow) corresponds to the absorption threshold of the *p*-type ZnO. Therefore, the near-band-edge emission at the *p-i-n* junction would be absorbed by the *p*-type overlayer even if it exists. It is also noteworthy that overall spectral line shape of the EL resembles that of the PL band observed for the *p*-type ZnO:N except for the spectral modulations, indicating that most of the radiative recombination occurs in the *p*-type layer. The three wavelengths pointed out above (420, 500 and 610 nm) satisfy Bragg's multiple reflection conditions $2d\sin\theta = m(\lambda/n)$, where $d = 620$ nm is the total ZnO layer thickness of the LED, $\theta$ equals $\pi/2$ for the vertical reflection discussed herein, *m* is positive integer, $\lambda/n$ is the wavelength in the material. Therefore, these three maxima are considered to be multiple internal reflection fringes of the EL band at 440 nm, which is presumably due to the DAP recombination.[15] Evidence to support this assignment are that the normalized spectral line shape of the EL was nearly independent of the forward current and ZnO layers are transparent for the broad EL band.

The unexpected observation in the EL spectra was the appearance of higher energy tail extending above the bandgap of ZnO ( > 3.3 eV), which was also found in the PL spectrum of the *p*-type ZnO:N, as shown in Fig. 3(b). The origin is not clear at present. However, one of the possible reasons would be a considerable downshift of the valence bands due to the repulsion with the localized N acceptor levels, which are broadened due to the heavy doping. The observation of the higher energy emissions means that electrons are injected deep into the *p*-type layer, since those emissions should be generated in the region close to the surface. Otherwise these emissions would be self-absorbed. The predominant radiative recombination in the *p*-type layer is due to its lower hole concentration discussed before, which gives rise to deep penetration of the depletion layer. To obtain the near-band-edge UV emissions, it is necessary to have higher hole concentration ( > $10^{18}$ cm$^{-3}$) in the *p*-type ZnO to turn on the hole injection into the undoped active regions. Formation of single- or double heterostructures containing *p*-type (Mg,Zn)O may also enable one to extract the UV lights efficiently.

Figure 4 shows the bias voltage (*V*) and the integrated EL intensity as a function of forward current bias (*I*). A good rectification was seen with a threshold voltage of 5.4 V, which is much lower than that reported previously (7 V).[1] This reduction in the threshold voltage might be due to the increased hole concentration in *p*-type ZnO, although any data of hole concentration for this particular or equivalent quality *p*-type ZnO layers are not available. We do not routinely characterize Hall effect of *p*-type ZnO layers now, because we think optimizing the growth conditions by characterizing device performances would be much easier and much more straight forward. In the forward bias region with *V* above the threshold, the current does not show exponential increase as *V* is increased, but is limited by an ohmic resistance of 380 Ω. This resistance value is in the same order as that measured between bottom electrodes of adjusted devices, indicating that current-spread resistance in *n*-type ZnO: Ga limits the current. This factor can be eliminated by employing heavily doped thicker lateral conducting layer (currently *n*-type ZnO doped with $10^{18}$ cm$^{-3}$ of Ga has a conductivity of $10^{-2}$ Ωcm with a carrier concentration of $10^{18}$ cm$^{-3}$ and a mobility of 100 cm$^2$V$^{-1}$s$^{-1}$). However, the use of conducting ZnO substrate (about 1 Ωcm is available) and backside contact of electrode will reduce the series resistance to 1 Ω and provide with much more simple device structure. Above *I* = 7 mA, the integrated EL intensity between 320 and 720 nm linearly increased with the current, as shown by the closed circles in Fig. 4. Although finite EL was observed at lower currents, the EL intensity is still weak and nonradiative recombination through the defects seems to dominate the power consumption. Heating effect seems to induce the saturation behavior at higher current region than 16 mA.

In summary, we have reported on the bluish EL spectra observed through the semi-transparent top electrode attached *p*-type ZnO of a *p-i-n* LED. The emission is assigned as a radiative recombination in the *p*-type ZnO, presumably that through DAP. It should be noted that broad yellow luminescence band centered at 2.2 eV observed in the previous LED[1] is completely eliminated in the present LED, making the color of the emission more bluish. The result clearly indicated the reduced point defect density in the present material. We identify near future challenges as follows: (1) improving the hole concentration above $10^{18}$ cm$^{-3}$ in *p*-type ZnO, (2) successful doping of (Mg,Zn)O into *p*-type to enable double-hetero structure, and (3) surface control of ZnO substrates to mimic the surface of annealed ZnO buffer layer on SCAM.

Acknowledgement

The authors thank to S. Fuke, H. Koinuma, T. Makino, M. Ohtani, Y. Segawa, and M. Sumiya for fruitful discussion and help for experiments. This work was supported by MEXT Grant (Creative Scientific Research 14GS0204 and 21st Century COE program "Promotion of Creative Interdisciplinary Materials Science for Novel Functions"), the Asahi Glass Foundation and the inter-university cooperative program of the IMR, Tohoku University. A. T. is supported by a JSPS fellowship.

15) A. Zeuner, H. Alves, D. M. Hofmann, B. K. Meyer, A. Hoffmann, U. Haboeck, M. Strassburg, and M. Dworzak, Phys. Stat. Sol. (b) **234**, R7 (2002).

**Figure captions**

Fig. 1 Schematic cross-sectional view of a ZnO *p-i-n* homojunction diode. Semi-transparent Au (5 nm) / Ni (5 nm) electrode is used for the contact to *p*-type ZnO layer.

Fig. 2 Optical microscope images of the device taken (a) under illumination and (b) in dark with feeding a direct forward bias current of 5 mA.

Fig. 3 (a) EL spectra at 293 K taken from the top of the device with feeding direct forward bias current of 8 mA, 12 mA and 16 mA. An arrow indicates the energy of sharp cut-off corresponding to the absorption edge of *p*-type ZnO top layer. (b) PL spectra at 293 K of undoped and *p*-type ZnO layers (dotted lines) and transmission spectrum (broken line) of *p*-type ZnO layer. Thickness of the undoped and *p*-type films are 1 μm and 500 nm, respectively.

Fig. 4 Current-voltage characteristics (solid line) of the ZnO *p-i-n* LED and forward bias current dependence of integrated EL intensity (closed circle).

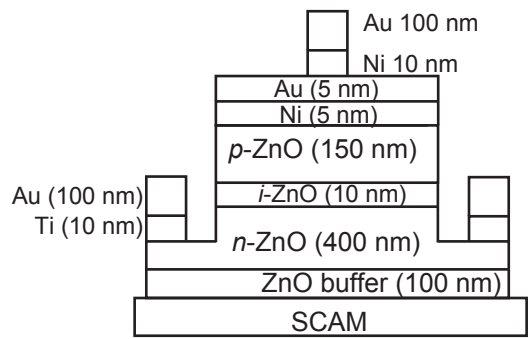

Figure 1  A. Tsukazaki et al.

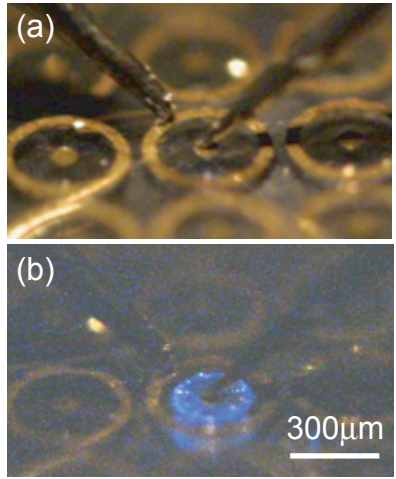

Figure 2  A. Tsukazaki et al.

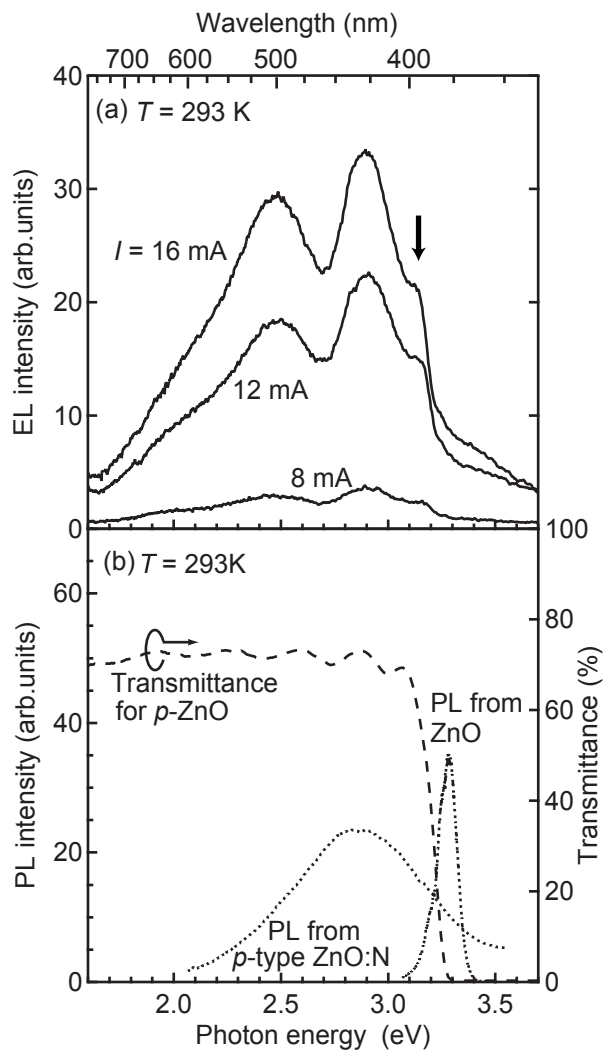

Figure 3  A. Tsukazaki et al.

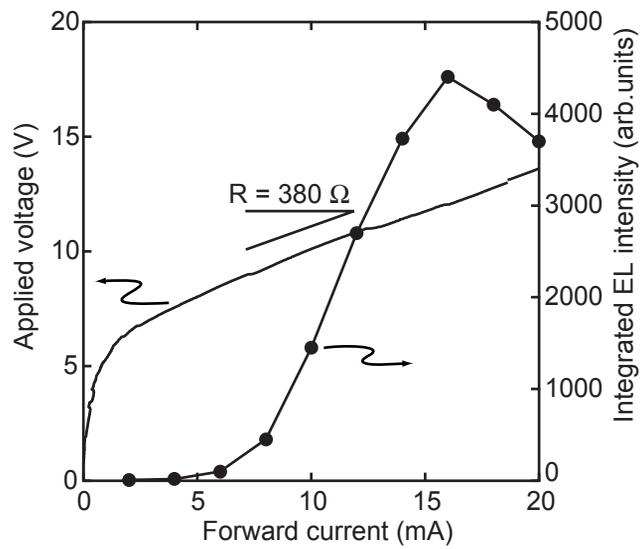

Figure 4  A. Tsukazaki et al.